\documentclass[letterpaper,10pt,twocolumn,final,journal,oneside]{IEEEtran}

\usepackage{cite}
\usepackage[Gray,amssymb]{SIunits}
\usepackage{color}
\usepackage{acronym}
\usepackage{esvect}
\usepackage[hidelinks]{hyperref}
\usepackage{amsfonts,amsmath,amssymb}
\usepackage{dsfont}
\usepackage{nicefrac}
\usepackage{lipsum}
\usepackage{multirow}
\usepackage{booktabs}
 \usepackage[table]{xcolor}
\ifCLASSINFOpdf
  \usepackage[pdftex]{graphicx}
  \DeclareGraphicsExtensions{.pdf,.jpeg,.png}
\else
  \usepackage[dvips]{graphicx}
  \DeclareGraphicsExtensions{.eps}
\fi
\usepackage{datetime}

\usepackage{ccicons}
\usepackage{tikz}

\definecolor{orange}{rgb}{1,0.5,0}

\newcommand{\R}{\mathbb{R}}
\newcommand{\T}{\mathrm{T}}
\renewcommand\vec[1]{\boldsymbol{#1}}%

\acrodef{hvdc}[\textsc{hvdc}]{high voltage direct current}
\acrodef{mtdc}[\textsc{mtdc}]{multi-terminal direct current}
\acrodef{pfs}[\textsc{pfs}]{power flow solution}
\acrodef{pcc}[\textsc{pcc}]{point of common coupling}
\acrodef{poc}[\textsc{poc}]{point of connection}
\acrodef{opf}[\textsc{opf}]{optimal power flow}
\acrodef{hb}[\textsc{hb}]{harmonic balance}
\acrodef{smps}[\textsc{smps}]{switched mode power supply}
\acrodef{mvf}[\textsc{mvf}]{multi-variate formulation}
\acrodefplural{smps}[\textsc{smps}]{switched mode power supplies}
\acrodef{pll}[\textsc{pll}]{phase locked loop}
\acrodef{sh}[\textsc{sh}]{shooting}
\acrodef{dae}[\textsc{dae}]{differential algebraic equation}
\acrodef{ode}[\textsc{ode}]{ordinary differential equation}
\acrodef{sde}[\textsc{sde}]{stochastic differential equation}
\acrodef{efm}[\textsc{efm}]{envelope-following method}
\acrodef{bvp}[\textsc{bvp}]{boundary value problem}
\acrodef{lte}[\textsc{lte}]{local truncation error}
\acrodef{lcm}[\textsc{lcm}]{least common multiple}
\acrodef{rmse}[\textsc{rmse}]{root mean squared error}
\acrodef{vc}[\textsc{vc}]{virtual connector}
\acrodef{ps}[\textsc{pss}]{power electromechanical sub-system}
\acrodef{ess}[\textsc{ess}]{electromagnetic sub-system}
\acrodef{mna}[\textsc{mna}]{modified nodal analysis}
\acrodef{psm}[\textsc{psm}]{power system model}
\acrodef{emt}[\textsc{emt}]{electromagnetic transient}
\acrodef{ts}[\textsc{ts}]{transient stability}
\acrodef{shpf}[\textsc{shpf}]{shooting power flow}
\acrodef{pf}[\textsc{pf}]{power flow}
\acrodef{vsc}[\textsc{vsc}]{voltage source converter}
\acrodef{mppt}[\textsc{mppt}]{maximum power point tracker}
\acrodef{pic}[\textsc{pi}]{proportional-integral}
\acrodef{ibr}[\textsc{ibr}]{inverter-based resource}
\acrodef{hvdc}[\textsc{hvdc}]{high-voltage direct current}
\acrodef{mtdc}[\textsc{mtdc}]{multi-terminal direct current}
\acrodef{iqr}[\textsc{iqr}]{interquartile range}
\acrodef{ou}[\textsc{ou}]{Ornstein-Uhlenbeck}
\acrodef{psd}[\textsc{psd}]{power spectral density}
\acrodef{coi}[\textsc{coi}]{center of inertia}
\acrodef{cig}[\textsc{cig}]{converter-interfaced generator}

\newcommand{\RefFig}[1]{Fig.~\ref{#1}}

\newcommand{\InRefEq}[1]{Equation~(\ref{#1})}

\newcommand{\RefE}[1]{(\ref{#1})}

\newcommand{\RefSec}[1]{Sec.~\ref{#1}}

\IEEEoverridecommandlockouts

\newcommand\copyrighttext{%
  \footnotesize
  \centering\ccby\\ \copyright~2023 the authors. This work is licensed under a \href{http://creativecommons.org/licenses/by/4.0/}{Creative Commons Attribution 4.0 International (CC BY 4.0 ) License}.\\
  IEEE Trans. on Power Sys.
  DOI:\href{https://doi.org/10.1109/TPWRS.2023.3236059}{10.1109/TPWRS.2023.3236059}}
\newcommand\copyrightnotice{%
\begin{tikzpicture}[remember picture,overlay]
\node[anchor=north,yshift=0pt] at (current page.north)
{\setlength{\fboxrule}{0pt}\fbox{\parbox{\dimexpr\textwidth-\fboxsep-\fboxrule\relax}{\copyrighttext}}};
\end{tikzpicture}%
}

\begin{document}

\title{Inertia Estimation Through Covariance Matrix}

\author{Federico Bizzarri, \IEEEmembership{Senior Member,~IEEE},
  Davide del Giudice, \IEEEmembership{Student Member,~IEEE}, Samuele
  Grillo, \IEEEmembership{Senior Member,~IEEE}, Daniele Linaro,
  \IEEEmembership{Member,~IEEE}, Angelo Brambilla,
  \IEEEmembership{Member,~IEEE}, and Federico Milano,
  \IEEEmembership{Fellow,~IEEE}\thanks{F.~Bizzarri is with Politecnico
    di Milano, DEIB, p.zza Leonardo da Vinci, no. 32, 20133 Milano,
    Italy and also with the Advanced Research Center on Electronic
    Systems for Information and Communication Technologies E.~De
    Castro (ARCES), University of Bologna, 41026 Bologna, Italy.
    (e-mail: federico.bizzarri@polimi.it).} \thanks{D.~del Giudice,
    S.~ Grillo, D.~Linaro, and A.~Brambilla are with Politecnico di
    Milano, DEIB, p.zza Leonardo da Vinci, no.~32, Milano, 20133,
    Italy.  (e-mails: \{davide.delgiudice, samuele.grillo,
    daniele.linaro, angelo.brambilla\}@polimi.it).} \thanks{F.~Milano
    is with School of Electrical \& Electronic Eng., University
    College Dublin, Belfield Campus, Dublin, D04V1W8, Ireland.
    (e-mail: federico.milano@ucd.ie).} \thanks{{Italian MIUR
      project PRIN 2017K4JZEE\_006 funded the work of S. Grillo
      (partially) and D. del Giudice (totally).}}}

\IEEEaftertitletext{\copyrightnotice\vspace{0.2\baselineskip}}
\maketitle

\begin{abstract}
  This work presents a technique to estimate on-line the inertia of a
  power system based on ambient measurements.  The proposed technique
  utilizes the covariance matrix of these measurements and solves an
  optimization problem that fits such measurements to the synchronous
  machine classical model.  We show that the proposed technique is
  adequate to accurately estimate the actual inertia of synchronous
  machines and also the virtual inertia provided by the controllers of
  converter-interfaced generators that emulate the behavior of
  synchronous machines.  {We also show that the proposed approach
    is able to estimate the equivalent damping of the classical
    synchronous machine model.  This feature is exploited to estimate
    the droop of grid-following converters, which has a similar effect
    of the swing equation equivalent damping.}  The technique is
  comprehensively tested on a modified version of the IEEE 39-bus
  system {as well as on a dynamic 1479-bus model of the
    all-island Irish transmission system}.
\end{abstract}


\vspace{-2mm}
\section{Introduction}

\subsection{Motivation}

The inertia constant has become a volatile parameter in power
systems with high penetration of renewable and non-synchronous
resources \cite{ULBIG20147290}.  Therefore, the estimation of the available
inertia is a valuable information that can help
system operators to ensure the security of the grid \cite{8450880}.
{This paper addresses this topic and aims at developing an on-line
inertia estimation method based on ambient measurements,
their stochastic behavior and a fair knowledge of the grid model}.

\vspace{-3mm}
\subsection{Literature Review}

There are mainly two methods for the on-line estimation of the inertia
available in a power system: (i) methods based on the measurement of
frequency and power variations after a disturbance or by injecting a
probing signal and (ii) methods that utilize ambient measurements
\cite{Tan2022,Dimoulias:2022}. Both methods have advantages and shortcomings.

The methods that are based on disturbances can be very accurate
\cite{Tamrakar2020} provided that the on-set of the disturbance is
correctly detected \cite{Wall2014, Zografos2018, Ashton2015,
delGiudice2019}. This can be challenging in low-inertia power
systems, which show a ``rich'' dynamical behavior \cite{Heylen2021}.
Moreover, the estimation cannot be performed continuously, but only
following disturbances \cite{8930036} \cite{8703783}.
Methods based on signal probing must inject active power of
adequate magnitude and require the installation of specific equipment
or the modification of pre-existing controllers~\cite{7913675}.

The methods that utilize ambient measurements are conceptually more
challenging than those based on disturbances.  The main advantage of
these methods is that they employ measurements obtained in normal
operating conditions and hence the estimation can be performed in a
continuous fashion.  The starting point is to build the dynamical
model of the system, which can be either complete or reduced to a
small number of coherent areas~\cite{8810634,Tuttelberg:2018}. Then,
the values of the parameters and of the observable state variables are
fit to available measurements. {It is worth mentioning that the
  number of areas and, consequently, the number of equivalent
  generators, whose inertia is to be estimated, is relevant.
  In~\cite{Gorbunov:2022}, the authors raise the question whether the
  \ac{coi} of a single-machine-equivalent power system model could be
  identified through ambient measurements. The answer given is that
  the combined effects of a well-damped \ac{coi} dominant mode and of
  a low-pass-filter action of the \ac{ou} processes modelling load
  behavior make it practically impossible to perform this
  task. However, this conclusion does not hold when a multi-machine
  system is considered.}

In {the latter group of inertia estimation methods}, we cite, for
example, \cite{Zhao2019}, where the authors obtain the fitting based
on a robust Kalman filter.  In~\cite{7913675} such a task is
performed using a closed-loop identification
method. In~\cite{9099871,baruzzi2021,baruzzi2022} inertia estimation
is performed by observing the step response of the reduced-order model
obtained from the full state-space model describing the grid, which
is, in turn, obtained through parameter identification using ambient
measurements. In~\cite{9123366} the power grid is reduced to a
two-machine system and then the system inertia is estimated using
inter-area oscillation modes. {In~\cite{Wang:2022}, the swing
  equation of the area single-machine-equivalent generator is
  transformed in Fourier domain. Then, the electromechanical
  oscillation, obtained by applying a fast Fourier transform to the
  generator estimated electrical power and rotor speed, are used to
  estimate the equivalent inertia. This method requires capturing the
  whole electromechanical oscillation trajectory.} The approach
proposed in \cite{Guo2022} estimates the inertia constants of each
generator by applying the regression theorem to a set of \ac{ou}
processes.  The method utilizes the covariance and correlation
matrices of the state variables to estimate the state matrix of the
system.  Then, the inertia constants are obtained solving a
least-square problem.  The main limitation of \cite{Guo2022} is the
assumption that the relationship between the active power variations
of synchronous machines and the voltage angle variations, as well as
the active and reactive power variations of converter-interfaced
generators are linear. {In a similar way, in~\cite{Guo:2022IEEE},
  the regression theorem of a multivariate \ac{ou} process is applied
  to power systems including \acp{cig}, modeled in detail with their
  \acp{pll} and supposed to work at unity power factor. The estimation
  is carried out separately for \acp{cig} and synchronous generators
  by identifying the relevant submatrices of the system state matrix.}

\vspace{-3mm}
\subsection{Contribution}

The approach proposed in this work falls in the group of methods that
use ambient measurements. We propose to exploit the \textit{colored
  noise} due to random fluctuations of the power consumption of the
loads and their impact on bus voltages, line currents, and power flows
across different areas of a power system. Then we define analytically
the statistical properties of the colored noise of these electrical
quantities as functions of the system parameters and of the inertia
present in the system. Finally, we utilize the variance of the
electrical measurements to minimize a non-linear least-squares cost
function with respect to the equivalent inertia constants as well as
the equivalent damping coefficients of the generators, either
synchronous {(see, e.g., equation (3.203) in \cite{Kundur:1994})}
or power-electronic based, that are connected to the grid.  The
solution is the sought value of the inertia constant or of the damping
coefficient that best fits the measurements and that satisfies the
grid constraints.

{The main benefit of the proposed approach is that it does not
  require a contingency or a large disturbance to occur to be able to
  estimate the inertia (or the damping). Only ambient measurements are
  needed, together with a fair knowledge of the model of the grid. The
  price for this is the need to take measurements in a given time
  period. Nevertheless, as shown in the case studies, the proposed
  approach is accurate and provides estimations of the inertia in the
  same time scale of short-term dispatch and adjustments markets.  The
  proposed method is original, as it shows for the first time how to
  perform inertia estimation by resorting to the covariance matrix of
  the power system without needing to estimate the state of the
  network or to derive (or measure) the rotor speed or any other
  internal variable of both synchronous machines and other devices
  performing frequency control.}

\vspace{-3mm}

\subsection{Organization}

The remainder of the paper is structured as follows.  Section
\ref{S:standard} gives the theoretical mathematical background of
power systems’ modeling and in Section \ref{S:NOISEINJ} the
mathematical model is enriched with the introduction of noise
injections to the power system, being this stage instrumental to model
load variability and perform the estimation. Section \ref{S:IE}
presents the proposed estimation procedure. Then the proposed method
is validated through numerical simulations, which are discussed in
Section \ref{S:NR}.  This section also discusses the challenges posed
by the proposed approach and presents a solution based on a proper
filtering of the measurements.  Finally, conclusions are drawn in
Section \ref{S:Con}.

\section{Power System Model}
\label{S:standard}

The proposed estimation procedure assumes that the generators
connected to the grid are synchronous machines and fits the
measurements to this model. In this section, we formulate the
equations of the system according to this assumption. It is important
to note, however, that every simulation is carried out utilizing a
full-fledged model representing in detail the dynamical behavior of
all generators, either synchronous or non-synchronous, and
their controllers.

To describe the dynamics of the \ac{psm}, we consider the
following set of \acp{dae}
\begin{equation}
  \label{E:FullSys2}
  \begin{aligned}
    {\vec{\dot \delta}} &= \Omega \left( \vec \omega -
    \omega_0 \right) \, , \\
    \vec M \vec{\dot \omega} &=
    \vec P_m(\vec \omega, \vec x, \vec y) -
    \vec P_e(\vec \delta, \vec x, \vec v, \vec \theta, \vec y)
    - \vec D (\vec \omega-\omega_0) \\
    \vec T \vec{\dot x}  &= \widehat{\vec F}\left(\vec \delta,
    \vec \omega, \vec x,  \vec v, \vec \theta, \vec y \right) \, , \\
    \vec 0 &=  {\vec G}\left(\vec \delta, \vec \omega, \vec x, \vec v,
    \vec \theta, \vec y \right)~,
  \end{aligned}
\end{equation}
where, assuming that $M$ is the number of machines, the meaning of
symbols in \RefE{E:FullSys2} is as follows:
\begin{itemize}
\item[--] $\Omega$: the base synchronous frequency in
  $\rad/\second$;
\item[--] $\vec \omega(t) \in \mathds{R}^{M}$: the per-unit rotor speeds of
  the machines;
\item[--] $\omega_0 \in \mathds{R}$: the per-unit reference synchronous
  frequency;
\item[--] $\vec \delta(t) \in \mathds{R}^{M}$: the rotor angles of
  the machines;
\item[--] ${\vec M \in \mathds{R}^{M \times M}}$: a diagonal matrix
  whose entries model the inertia constants ${H}$ of the
  machines. {In particular, $M_{jj} = 2H_{jj}$ (for
    ${j=1,\dots,M}$);}
\item[--] $\vec x(t) \in \mathds{R}^{N}$: the $N$ state variables of
  the \ac{psm} ($\vec \omega$ and $\vec \delta$ excluded) that can
  influence the dynamics of the machines, and
  $\vec T \in \mathds{R}^{N \times N}$ is a mass matrix;
\item[--] $\vec y(t) \in \mathds{R}^{S}$: all the algebraic
  variables of the \ac{psm} but $\vec v$ and $\vec \theta$;
\item[--] ${\vec D \in \mathds{R}^{M \times M}}$: a diagonal matrix
  whose entries $d_{jj}$ (for ${j=1,\dots,M}$) model the damping
  factor of the machines;
\item[--] ${\vec v(t) \in \mathds{R}^{P}}$ and
  ${\vec \theta(t) \in \mathds{R}^{P}}$: bus voltages and phases,
  respectively, where $P$ is the number of buses;
\item[--] ${\vec P_m(\vec \omega, \vec x, \vec y) \in \mathds{R}^{M}}$:
  the mechanical power regulated by controllers depending on
  $\vec \omega$, $\vec x$, and $\vec y$;
\item[--]
  $\vec P_e(\vec \delta, \vec x, \vec v, \vec \theta, \vec y) \in
  \mathds{R}^{M}$: the electrical power exchanged by machines;
\item[--] $\widehat{\vec F}(\cdot)$ accounts for regulators and other
  dynamics included in the system; and
\item[--]${\vec G}(\cdot)$ models algebraic constraints such as lumped
  models of transmission lines, transformers and static loads.
\end{itemize}

\InRefEq{E:FullSys2} can be conveniently rewritten as
\begin{equation}
  \begin{aligned}
    \vec\Lambda \dot{\vec \xi} &=
    \widetilde {\vec F}(\vec \xi, \vec \zeta) \, , \\
    \vec 0 &= \vec G(\vec \xi, \vec \zeta)~,
  \end{aligned}
  \label{E:FullSys3}
\end{equation}
where
$\displaystyle{\vec \xi(t) \equiv [\vec \omega, \vec \delta, \vec
  x]^\T}$,
$\displaystyle{\vec \zeta(t) \equiv [\vec v, \vec \theta, \vec
  y]^\T}$, and $\vec \Lambda$ is a non-singular diagonal matrix
	{including $\vec M$}.
$\vec G$ and $\widetilde {\vec F}$ are assumed to be continuously
differentiable in their definition domain and matrices of their
partial derivatives are referred to as $\widetilde {\vec F}_\xi$,
$\widetilde {\vec F}_\zeta$, $\vec G_\xi$, and $\vec G_\zeta$.  As an
example,
$\widetilde {\vec F}_{\xi_{jk}} = {\partial \widetilde {\vec
    F}_j}/{\partial \xi_k}$, for $j,k =1,...,2M+N$.

The implicit function theorem guarantees that if
$\vec G(\vec \xi^*,\vec \zeta^*)= \vec 0$, provided that
$\vec G_{\zeta}(\vec \xi^*,\vec \zeta^*)$ is not singular, a unique
and smooth function $\vec\Gamma: \R^{2M+N} \rightarrow \R^{2P+S}$
exists so that $\vec \zeta^*=\vec \Gamma(\vec \xi^*)$.  If the
conditions of the implicit function theorem are satisfied,
\RefE{E:FullSys3} can be rewritten as
\begin{equation}
  \begin{aligned}
    \vec \Lambda \dot{\vec \xi}
    &=
      \widetilde {\vec F}(\vec \xi, \vec \Gamma(\vec \xi))
      \equiv \vec F(\vec \xi)~,
  \end{aligned}
  \label{E:Fullsys4}
\end{equation}
the equilibrium points of which, say $\vec \xi_o$, satisfy the
condition
\begin{equation}
  \vec F(\vec \xi_o) =  \vec 0~,
\end{equation}
with the Jacobian matrix
\begin{equation}\!
  \vec J(\vec \xi_o) \! = \! \vec \Lambda^{-1} \! \vec F_\xi \! = \!
  \vec \Lambda^{-1}(\widetilde {\vec F}_\xi \! - \! \left.
  \widetilde {\vec F}_\zeta \vec G_\zeta^{-1} \! \vec G_\xi)\right|_{\vec \xi= \vec \xi_o,
	\vec \zeta = \vec \Gamma(\vec \xi_o)}~.
  \label{E:JFullsys4}
\end{equation}

\section{Inclusion of Noise}
\label{S:NOISEINJ}

We assume stochastic noise is injected into the \ac{psm} as
\begin{equation}
  \begin{aligned}
    \vec\Lambda \dot{\vec \xi} &=
    \widetilde {\vec F}(\vec \xi, \vec \zeta) \, , \\
    \vec 0 &= \vec G(\vec \xi, \vec \zeta) + \vec \Xi \, \vec \eta  ~,
  \end{aligned}
  \label{E:Noise1}
\end{equation}
where $\vec \eta$ is a vector of $Z$ independent \acf{ou} processes
\cite{Arnold1974}, and $\vec \Xi\in\R^{(2P+S) \times Z}$ is a constant
matrix.

The \ac{ou} processes are defined through a set of \acp{sde}, as follows
\begin{equation}
  \label{E:eta}
  d \vec \eta = -\vec \Upsilon \vec \eta \, dt +
  \vec \Sigma \, d \vec W_t ~,
\end{equation}
where the drift $\vec \Upsilon \in \R^{Z \times Z}$ and diffusion
$\vec \Sigma\in \R^{Z \times Z}$ are diagonal matrices with positive
entries, $\vec W_t \in \R^Z$ is a vector of $Z$ Wiener processes, and
the differentials rather than time derivatives are utilized to account
for the idiosyncrasies of the Wiener processes.
The \ac{ou} processes are characterized by a mean-reversion property and
show bound standard deviation.  Moreover, these processes show a
spectrum that is an accurate model of the stochastic variability
of power loads \cite{HIRPARA2015409, NWANKPA1990338, Milano2013,
  7540898, 8447491}.  In \eqref{E:Noise1}, we assume that noise
injection models the effect of the consumption randomness of $Z$ loads.

Note that, in \eqref{E:Noise1}, the vector of stochastic processes
$\vec \eta$ appears only in the algebraic equations.  This is assumed
for simplicity but without lack of generality.  The interested reader
can find for instance in \cite{Milano2013} a stochastic \ac{psm} model where
the noise perturbs both differential and algebraic equations.

We also assume that the action of $\vec \eta$ can be safely modeled as
a small-signal perturbation around an equilibrium point.  Hence, the
solution of \RefE{E:Noise1} can be written as
$\vec \xi=\vec \xi_{o} + \vec \xi_\eta$ and
$\vec \zeta=\vec \zeta_{o} + \vec \zeta_\eta$, that is, the effects of
perturbations are assumed as additive.  It is thus possible to
linearise \RefE{E:Noise1} around $\vec \xi_{o}$ and obtain the random
ordinary differential equation \cite{RODEBOOK}
\begin{equation}
  \begin{aligned}
    \vec \Lambda \dot{\vec \xi}_{\eta}
    &= \vec F_\xi
      \vec \xi_\eta - \widetilde{\vec F}_\zeta
      \vec G_\zeta^{-1} \vec \Xi \, \vec \eta ~,
  \end{aligned}
  \label{E:Noise2}
\end{equation}
where the ${\vec F}_\xi$, $\widetilde{\vec F}_\zeta$, and
${\vec G}_\zeta$ matrices are computed at $(\vec \xi_{o},\vec \zeta_{o})$.
The small-signal variations of the algebraic variables are given
by
\begin{equation}
  \vec \zeta_\eta = \underbrace{-\vec G_\zeta^{-1}\left[\,\vec G_\xi |
      \, \vec \Xi\,\right]}_{\vec E}\left[\!\!
    \begin{array}{c}
      \vec {\vec \xi_\eta}\\
      \vec {\vec \eta}
    \end{array}
    \!\!\right] ~.
  \label{E:Noise3}
\end{equation}
Augmenting \eqref{E:Noise2} with \eqref{E:eta} allows obtaining the linear
\acp{sde} (in narrow sense \cite{Arnold1974}) governing the overall
dynamics of the linearised noisy \ac{psm}.  Adopting the typical
formalism of the \acp{sde}, the complete set of linearised \acp{sde} reads
\begin{equation}\!
  d\!\!\underbrace{\left[\!\!\!
      \begin{array}{c}
        \vec {\vec \xi_\eta}\\
        \vec {\vec \eta}
      \end{array}
      \!\!\! \right]}_{\vec X_t}\!\!=\!\!
  \underbrace{\left[\!\!\!
      \begin{array}{cc}
        \overbrace{\vec \Lambda^{-1} {\vec F}_\xi}^{\vec J}
        \!\! & \!\! - \vec \Lambda^{-1} \widetilde{\vec F}_\zeta
          {\vec G}_\zeta^{-1} \vec \Xi\\
        \vec 0 \!\! & \!\! - \vec \Upsilon
      \end{array}
      \!\!\!\right]}_{\vec A}\!\!
  \left[\!\!\!
    \begin{array}{c}
      \vec {\vec \xi_\eta}\\
      \vec {\vec \eta}
    \end{array}
    \!\!\! \right]\!\!dt +
  \underbrace{\left[\!\!\!
      \begin{array}{c}
        \vec 0\\
        \vec \Sigma
      \end{array}
      \!\!\! \right]}_{\vec B}\!\!d\vec W_t ~,
  \label{E:Noise4}
\end{equation}
The solution of \RefE{E:Noise4} with a normally distributed (or
constant) initial condition is a $(2M+N+Z)-$dimensional Gaussian
stochastic process.

\section{Proposed Technique to Estimate the Inertia}
\label{S:IE}

The mean and the covariance matrix of process \RefE{E:Noise4} can be
derived by solving two sets of linear \acp{ode} \cite{Arnold1974}.
Since \RefE{E:Noise4} is stable under the hypothesis that
\RefE{E:Fullsys4} is stable at $\vec \xi_{o}$, these \acp{ode} reveal
that, at steady state, the mean of $\vec X_t$ is zero, and its
$\vec K_{X_t}$ covariance matrix derives from the solution of the
following Lyapunov equation
\begin{equation}
  \vec A \vec K_{X_t} + \vec K_{X_t} \vec A^\T +
  \vec B \vec B^\T = \vec 0 ~.
  \label{E:Noise5}
\end{equation}
The diagonal elements of $\vec K_{X_t}$ are the (steady-state)
variances of the components of the $\vec X_t$ process.
In particular, the last $Z$ diagonal elements, {corresponding
to the $\vec \eta$ sub-vector of $\vec X_t$, are the (steady-state) variances
of the $Z$ independent \ac{ou} processes introduced in \RefE{E:eta}. Hence,
these terms can be written as
$\sigma^{2}_{z}/ (2 \upsilon_{z})$, for $z=1,...,Z$, where $\sigma_z$
and $\upsilon_z$ are the diagonal elements of $\vec \Sigma$ and
$\vec \Upsilon$, respectively \cite{Arnold1974}. The remaining $2M+N$ diagonal
elements of $\vec K_{X_t}$ refer to the ${\vec \xi_\eta}$ sub-vector of $\vec X_t$
and are influenced by $\vec \eta$ through the sub-matrix $- \vec \Lambda^{-1} \widetilde{\vec F}_\zeta {\vec G}_\zeta^{-1} \vec \Xi$ in \RefE{E:Noise4}}.

{According to \RefE{E:Noise3}, $\vec \zeta_\eta$ can be written as a
linear combination of the entries of $\vec X_t$ through $\vec E$, thus being
a multidimensional Gaussian process, too. Hence, it is possible to write the
$\vec K_{\zeta_\eta}$ covariance matrix of the
small-signal algebraic variables as the quadratic form \cite{provost1992,9968165}}
\begin{equation}
  \vec K_{\zeta_\eta} = \vec E \vec K_{X_t} \vec E^\T ~.
  \label{E:Noise6}
\end{equation}

The proposed approach exploits \eqref{E:Noise6} and the fact that
$\vec K_{\zeta_\eta} $ (and, thus, the variance of the algebraic
variables) depends through $\vec K_{X_t}$ and $\vec A$ on the
elements of $\vec \Lambda$, a subset of which are the inertia
constants of the synchronous machines.

Let us assume one wants to estimate a subset $\mathcal M$ of
these inertia constants and let $\mathcal{Z}$ be a set of \textsc{pmu} measurements
of bus voltages and line currents flowing through a given number of
transmission lines and/or transformers.

\begin{figure}[t]
  \centering
  \includegraphics[width=0.75\columnwidth]{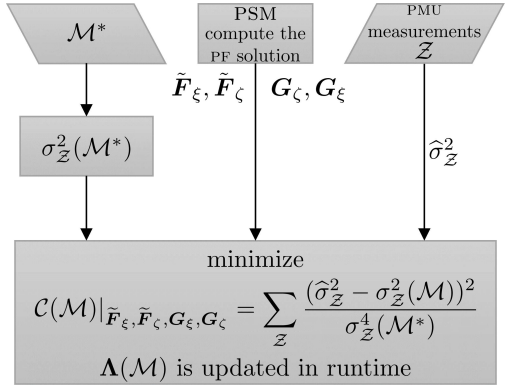}
  \caption{The flow-diagram of the proposed inertia estimation
    procedure.}
  \label{F:DIAGRAM}
\end{figure}

The above derivations show that, if the
model and the parameters of the grid are known, one can
compute the $\sigma^2_{\mathcal{Z}}$ variances of the measured
quantities, which are the diagonal elements of the
$\vec K_{\zeta_\eta}$ matrix.
However, it is also possible to solve an inverse problem:
knowing through measurements the variance of the elements of
$\mathcal Z$, the values of the elements of
$\mathcal M$ can be determined by minimizing the following non-linear least-squares cost
function (see also
Fig.~\ref{F:DIAGRAM})
\begin{equation}
  \left.
    \mathcal C(\mathcal{M})
  \right|_{\widetilde{\vec F}_\xi, \widetilde{\vec F}_\zeta, {\vec G}_\xi, {\vec G}_\zeta} =
  \sum_{\mathcal Z}{\frac{(\widehat \sigma^2_{\mathcal{Z}} - \sigma^2_{\mathcal{Z}}(\mathcal{M}))^2}{\sigma^4_{\mathcal{Z}}(\mathcal{M}^*)}}~,
\label{E:CF}
\end{equation}
where $\widehat \sigma^2_{\mathcal{Z}}$ are the variances of the
measured variables, $\sigma^2_{\mathcal{Z}}(\mathcal{M})$ are the
variances computed in run-time during the minimization procedure
w.r.t. the elements of $\mathcal{M}$, while
$\sigma^2_{\mathcal{Z}}(\mathcal{M}^*)$ are the corresponding
variances of the same variables obtained with the model for a
reference set of $\mathcal M$ values, which are used to normalize the
cost function. 
During the optimization process the estimation of the elements of $\mathcal M$ makes necessary to update, iteratively, $\vec \Lambda$ and, consequently, $\vec A$\footnote{Actually if, for instance, some specific
      implementation of virtual synchronous generators are considered,
      it may happen that the equivalent inertia constant of those
      devices directly affects some of the entries of the
      $ \vec F_{\xi}$ Jacobian matrix instead of $\vec \Lambda$.  This
      situation, despite being more involved, can be handled by
      updating $ \vec A$ during the optimization process and, for the
      sake of simplicity, we will not dwell further on this point
      here.}.

We note that the procedure above is \textit{agnostic} w.r.t. to the
devices that are connected to the grid. That is, the estimation of
the inertia is obtained by fitting the synchronous machine model to
the measurements. If there exist non-synchronous devices, such as
converter-interfaced generation, their ``equivalent'' inertia
constants can be obtained using \eqref{E:CF} regardless of the fact that
the converter is set up to be grid forming or grid following.
 The procedure is also agnostic in terms of the parameters to be
estimated, provided that such parameters do have an effect on the
frequency variations of the system. In particular, we use this
property to estimate also the equivalent damping coefficients of the
devices, which can be attained by minimizing a cost
function for a subset of damping coefficients $\mathcal{D}$,
in a similar way as for $\mathcal{M}$ in \eqref{E:CF},
\begin{equation}
  \label{eq:Dest}
  \left.
    \mathcal C(\mathcal{D})
  \right|_{\widetilde{\vec F}_\xi, \widetilde{\vec F}_\zeta, {\vec G}_\xi, {\vec G}_\zeta} =
  \sum_{\mathcal Z}{\frac{(\widehat \sigma^2_{\mathcal{Z}} - \sigma^2_{\mathcal{Z}}(\mathcal{D}))^2}{\sigma^4_{\mathcal{Z}}(\mathcal{D}^*)}}~.
\end{equation}

\section{Case Studies}
\label{S:NR}

{In this Section, we consider two grids, the well known IEEE
  39-bus system and a 1479-bus dynamic model of the all-island Irish
  transmission system \textsc{aiits}.
	The IEEE 39-bus system allows us illustrating
  the various features and challenges of the proposed method.  With
  this aim we first consider the conventional system with synchronous
  machines and their regulators (Section \ref{ss:1}).  Then, we
  evaluate the behavior of the proposed method when the system
  includes grid-forming (Section \ref{ss:2}) and grid-following
  converters (Section \ref{ss:3}).  The all-island Irish system, on
  the other hand, serves to demonstrate the robustness of the
  proposed method when applied to a real-world complex network.}

\subsection{IEEE 39-bus System}

\begin{figure}[t]
  \centering
  \includegraphics[width = \columnwidth]{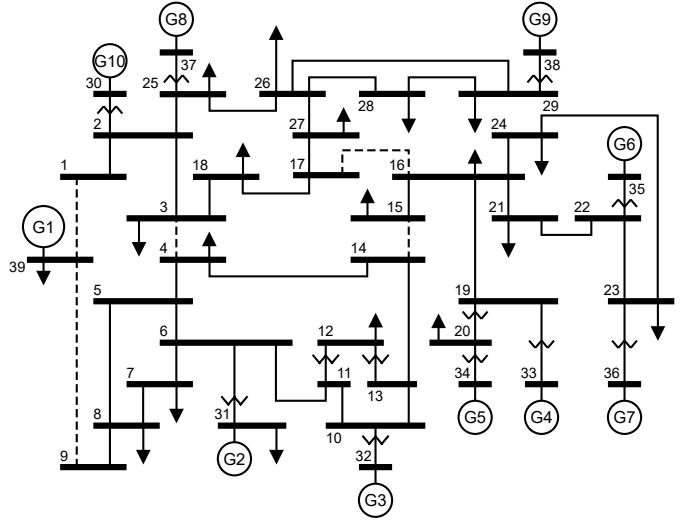}
  \caption{The single-line diagram of the IEEE 39-bus power system.}
  \label{F:IEEE39}
\end{figure}

We use as a benchmark the IEEE 39-bus power system shown in
\RefFig{F:IEEE39} \cite{4113518}.  This is a simplified model of the
transmission system in the New England region in the Northeastern
United States and is composed of $10$ generators, $34$ lines, $19$
loads, and $12$ transformers.  The network models and parameters are
those adopted and directly extracted from its DIgSILENT PowerFactory
implementation \cite{IEEE39a} and derived from \cite{Pai1989}.  The
$G_{1}$ generator, which represents the connection of the New England
System to the rest of the US and Canadian grid, is modeled with a
constant excitation \cite{Pai1989}, viz.~there is no automatic voltage
regulator (\textsc{avr}) connected to it.  On the contrary, the other
generators are equipped with \textsc{avr}s (specifically, rotating
excitation systems of IEEE Type 1 according to \cite{Pai1989}).
Governors are considered as IEEE Type G1 (steam turbine) for
$G_{2}$--$G_{9}$, and IEEE Type G3 (hydro turbine) for $G_{10}$.
Finally, in Table \ref{T:PRATH} we report the value of inertia
constants and power rating of the 10 generators.

\begin{table}[b]
  \caption{Synchronous generators $H$ and $P_{\mathrm{RAT}}$}
  \centering
      \begin{tabular}{ c | c | c | c | c | c}
      \toprule
      Gen. & $H\,[\second]$& $P_{\mathrm{RAT}}\,[\mega\watt]$ &
      Gen.	& $H\,[\second]$& $P_{\mathrm{RAT}}\,[\mega\watt]$\\
      \midrule
      $G_{1}$ & $5.00$  & $10000$ & $G_{6}$ & $4.35$ & $800$ \\
      $G_{2}$ & $4.33$  & $700$ & $G_{7}$ & $3.77$ & $700$\\
      $G_{3}$ & $4.47$  & $800$ & $G_{8}$ & $3.47$ & $700$\\
      $G_{4}$& $3.57$& $800$& $G_{9}$& $3.45$& $1000$\\
      $G_{5}$& $4.33$& $600$& $G_{10}$& $4.20$& $1000$\\
      \bottomrule
    \end{tabular}
    \label{T:PRATH}
\end{table}

The active power consumption of each of the $Z=19$ loads is altered by
one of the $\eta_z(t)$ \ac{ou} processes ($z=1,\dots,Z$) in
\eqref{E:Noise4}, which has zero mean, and variance
$\sigma_z^2/(2\upsilon_z)$.  All loads are assumed to incorporate
stochastic power fluctuations \cite{Milano2013}
\begin{equation}
  \mathrm{L}_z(t) = \eta_z(t) \, P_{{\mathrm{L0}}_z} \left(
  \frac{\left|V_z(t)\right|}{V_{0_z}} \right)^{\gamma}~,
\end{equation}
where (for $z=1,\dots,Z$) $P_{{\mathrm{L0}}_z}$ is the nominal active
power of the load, $V_{0_z}$ is the load voltage rating, $V_z(t)$ is
the bus voltage at which the load is connected and $\gamma$ governs
the dependence of the load on bus voltage. In the simulation below, we
assume $\upsilon_z = 0.5$ and $\sigma_z$ is defined in such a way that
the standard deviation of $\eta_z(t)$ is $5\%$ of $P_{\mathrm{L0}}$.
The zero mean implies that stochastic load power fluctuations do not
perturb, on average, the operating point of the system. Furthermore,
we assumed $\gamma=0$.

To carry out the simulations discussed below, the numerical
integration of the multi-dimensional \ac{ou} process in
\RefE{E:Noise1} was based on the numerical scheme proposed by
Gillespie in \cite{Gillespie1996}.  Furthermore, the second-order
trapezoidal implicit weak scheme for stochastic differential equations
with colored noise, available in the simulator \textsc{pan}
\cite{ngcas, Bizzarri201451,Linaro2022}, was adopted \cite{Milshtein1994}.

\subsubsection{Conventional Power System}
\label{ss:1}

The objective of the first scenario considered in this case study is
to estimate the inertia constants of the 10 generators of the IEEE
39-bus system across a time period spanning one day.  The target is to
obtain an estimation of the inertia constants every $15\,\minute$
(estimation window) of the working day.  To this end, we compute the
variance of the currents flowing through the generators, using a
$15\,\minute$-long moving window.  The values of variance are the
elements of the $\mathcal M$ set introduced in \RefSec{S:IE}.  The
sampling period is $25\,\milli\second$, i.e., a sampling rate of
$40\,\hertz$.\footnote{{The steady-state variance of a given
    time-domain stochastic signal corresponds to the integral of its
    \ac{psd} in the frequency domain. When computing the variance of a
    given set of samples of such a signal, one must choose a sampling
    frequency and a sampling window such that the corresponding
    \ac{psd} is accurate enough w.r.t. the continuous-time original
    signal. By doing so, the steady-state variance computed in time
    domain from a finite set of samples will be accurate
    enough. According to several numerical tests, we derived that a
    sampling period of $25\,\milli\second$ in a sampling window of
    $15\,\minute$ is a good tradeoff between accuracy and
    computational burden.}}

We also assume that the matrices $\tilde{\vec F}_{\vec \xi}$,
$\tilde{\vec F}_{\vec \zeta}$, ${\vec G}_{\vec \xi}$,
${\vec G}_{\vec \zeta}$ are constant, i.e., that the parameters of the
IEEE 39-bus system do not vary during the simulation.  This assumption
simplifies the analysis but is not a binding requirement of the
proposed procedure.  If the operating point and/or the topology of the
grid change, one just needs to update those matrices.

\begin{figure}[t]
  \centering
  \includegraphics[scale=1.0]{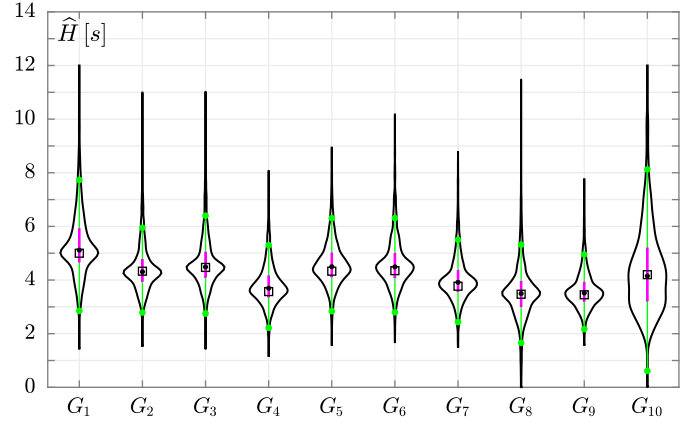}
  \caption{Violin plots of the inertia estimations of each generator
    of the IEEE 39-bus system. The black solid circle markers
    correspond to the median value of the optimization results,
    whereas the empty square markers represent the nominal values of
    the inertia constants. The magenta bars represent the \ac{iqr},
    viz. the spread difference between the $75^{\mathrm{th}}$ and
    $25^{\mathrm{th}}$ percentiles of the data. The green solid circle
    markers represent the upper adjacent value (i.e., the largest
    observation that is less than or equal to the third quartile plus
    $1.5 \times \ac{iqr}$) and the lower adjacent value (i.e., the
    smallest observation that is greater than or equal to the first
    quartile minus $1.5 \times \ac{iqr}$).  Time window: $15$
    min. Optimization performed on the unfiltered currents.}
  \label{F:SCE1a}
\end{figure}

The $\widehat H$ inertia constant estimates were obtained with the
MATLAB Global Optimization Toolbox and the {\tt fgoalattain} function.
The search interval of the optimization procedure was lower-bounded to
zero since inertia constants cannot be negative.  We performed $20$
independent trials each one being a $24\,\hour$ simulation of the grid
and an optimization problem is solved every $15\,\minute$ of simulated
time, thus collecting $1920$ estimates per synchronous generator. The
violin plots\footnote{A violin plot is a combination of a box plot and
  a kernel density plot.  Specifically, it starts with a box plot and
  then adds a rotated kernel density plot to each side of the box
  plot.} in {\RefFig{F:SCE1a}} (one for each generator) summarize
the obtained results.  The median of the estimated value of the
inertia constants (black solid dots in \RefFig{F:SCE1a}) are in good
agreement with the corresponding nominal values (empty squared markers
in \RefFig{F:SCE1a}) listed in Table \ref{T:PRATH}.  Despite the
amplitude of the \ac{iqr} intervals (magenta vertical bars in
\RefFig{F:SCE1a}) are reasonably narrow, the upper and lower adjacent
values (green solid dots in \RefFig{F:SCE1a}) are quite far from the
median values, and the same holds for the tails of the distribution of
the estimated values.  {The maximum value of the relative
  percentage error, computed as
  $\epsilon_{\%}^H = 100\cdot\frac{|\text{Median}(\widehat H) -
    H|}{H}$, is obtained for $G_5$ and amounts to $3.87\%$. The
  $0.44\%$ minimum value is achieved for $G_3$.}

\begin{figure}[htb]
  \centering
  \includegraphics[scale=1.0]{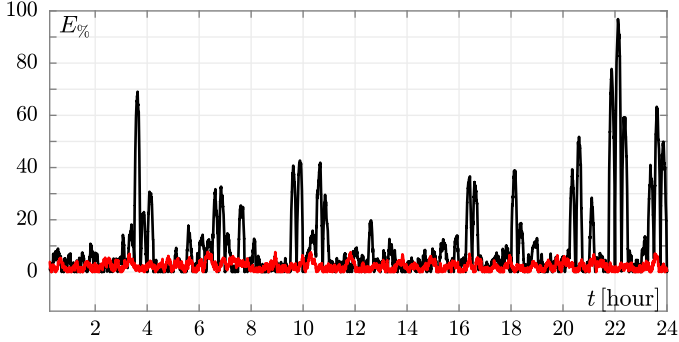}
  \caption{{The $E_\%$ percentage relative error} of the moving variance of the
    $i_{8}$ current, calculated over a sliding window of
    $15\,\minute$, with respect to the steady-state variance of
    $i_{8}$.  The black and the red curves refer to the unfiltered and
    filtered case, respectively.}
  \label{F:FIL_vs_NOFIL}
\end{figure}

The quality of the results can be improved by filtering the considered
currents before computing their moving variance.  This is done by
resorting to a steep bandpass filter acting from $0.1\,\hertz$ to
$1.5\,\hertz$.  This bandwidth was chosen to remove the low frequency
contribution of the \ac{ou} process.  This is done since these
components account for the very slow fluctuations of the currents that
heavily affect the trend of their moving variance w.r.t. the
steady-state values of the variance itself.  The effect of this filter
is shown in \RefFig{F:FIL_vs_NOFIL} for the $i_{8}$ current (i.e., the
current flowing through $G_8$) in terms of the absolute value of the
relative percentage error between the moving variance and its
steady-state value over $24\,\hour$ in one of the $20$ trials.  In
particular, the black and red curves refer to the unfiltered and the
filtered case, respectively, and the improvement over the unfiltered
case is evident.

\begin{figure}[htb]
  \centering
  \includegraphics[scale=1.0]{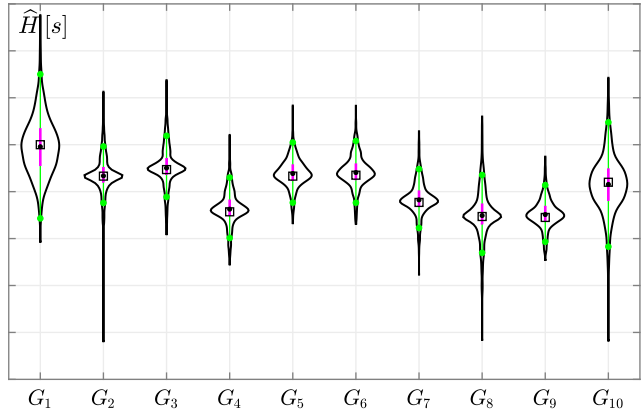}
  \caption{Violin plots of the inertia estimations of each generator
    of the IEEE 39-bus system. The optimization was performed on the
    filtered currents, on a time window of $15\,\minute$. Note the
    different scale of the vertical axes w.r.t. \RefFig{F:SCE1a}.}
  \label{F:SCE1b}
\end{figure}

The state equations governing the dynamics of the filters were added
to the set of \acp{ode} reported in \RefE{E:Fullsys4} and this allowed
to derive the steady-state variance of the filters output solving
\RefE{E:Noise5}. The inertia constants of the IEEE 39-bus generators
was estimated from the filtered currents and the results are reported
in \RefFig{F:SCE1b}.  The effect of this signal processing is a
significant reduction of the amplitude of the interval spanned by the
upper and lower adjacent values, thus guaranteeing a more reliable
estimate of the inertia constants.  {The $\epsilon_{\%}^H $
  relative percentage error spans from $0.14\%$ (for $G_2$) to
  $1.78\%$ (for $G_9$).}  These results illustrate well that the
proposed approach accurately estimates the inertia constants of the
ten synchronous generators of the system.

\subsubsection{Inclusion of Grid-Forming Converters}
\label{ss:2}

This section illustrates the behavior of the proposed estimation
technique for a scenario where the IEEE 39-bus system is modified to
include a variable generation/load profile and grid-forming
converters.  These devices are assumed to mimic the behavior of
synchronous machines and are thus expected to provide a
\textit{virtual} inertia to the system.  {The controllers of
  grid-forming converters are based on \cite{9625933}.}  The objective
of this section is twofold: (i) showing that the proposed inertia
estimation approach works correctly under variable generation/load
conditions and (ii) illustrating the ability of the proposed
estimation technique to capture the effect of grid-forming converters.

\begin{figure}[t]
  \centering
  \includegraphics[scale=1.0]{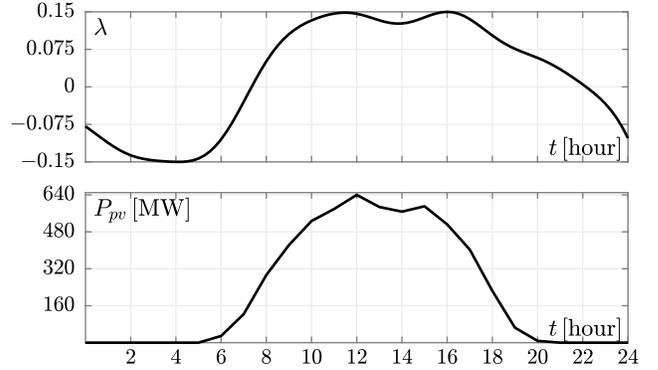}
  \caption{Upper panel: time evolution of the $\lambda$ coefficient
    used to overload the IEEE 39-bus system. Lower panel: the overall
    active power supplied by the solar plants that were added to the
    IEEE 39-bus system.}
  \label{F:DUCK_SOLAR}
\end{figure}

For what regards point (i), we first modified the IEEE 39-bus system
by mimicking a typical daily variation of the loads. To do so, we
multiplied the power absorbed by the loads and the (active) power
generated by the synchronous machines by a coefficient $\lambda$. To
alter the nominal value of the power loads, $\lambda$ was varied
continuously in time according to the profile shown in the upper panel
of \RefFig{F:DUCK_SOLAR}. This profile was sampled every
$15\,\minute$: this same time basis was used to update the set points
of the generators. The effect of energy production by three
(aggregated) solar plants connected to $\tt{bus_{8}}$,
$\tt{bus_{24}}$, and $\tt{bus_{27}}$ was emulated by reducing the
absorbed active power of the loads connected at those buses. Load
power was decreased by the same active power supplied by those
(aggregated) solar plants (each one supplying one third of the
electrical power reported in the lower panel of
\RefFig{F:DUCK_SOLAR}). To balance generation, the power supplied by
these plants was subtracted from the power of synchronous generators.
In particular, the active power supplied by $G_3$, $G_7$, and $G_8$
was varied in proportion to their power ratings.
% We recall that the instantaneous mechanical power by the prime
% movers is regulated by turbine governors.
Concerning point (ii), the power rating of each one of the virtual
synchronous generators (i.e., the solar plants connected to
$\tt{bus_{8}}$, $\tt{bus_{24}}$, and $\tt{bus_{27}}$) is
$100\,\mega\watt$.  We regulated these generators so that each of them
provided, through the controllers of their inverters, a total virtual
inertia equivalent to $H=10\,\second$ only from 8:15~\amname{} to
5:30~\pmname{}.

\begin{figure}[t]
  \centering
  \includegraphics[scale=1.0]{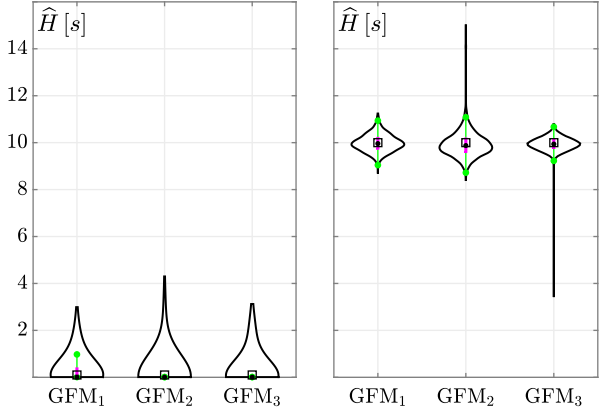}
  \caption{Violin plots of the inertia constants estimates of the
    grid-forming converters $\rm GFM_1$, $\rm GFM_2$, and $\rm GFM_3$
    (all of which were set to $10\,\second$ from 8:15~\amname{} to
    5:30~\pmname{}).  Left and right panels refer to the low- and
    high-inertia cases, respectively. The time window is
    $15\,\minute$.}
  \label{F:SCEC}
\end{figure}

Results are shown in \RefFig{F:SCEC}, using again violin plots. The
left panel refers to the time interval in which the solar plants are
not connected and thus the inertia constants of their grid-forming
converters is zero.  The right panel corresponds to the 8:15~\amname{}
-- 5:30~\pmname{} time interval.  {In the high-inertia case, the
  $\epsilon_{\%}^H$ percentage error for $\rm GFM_1$, $\rm GFM_2$, and
  $\rm GFM_3$ is equal to $0.29\%$, $1.23\%$, and $0.60\%$,
  respectively. In the low-inertia case, $\epsilon_{\%}^H$ is almost
  the same for the three grid-forming converters and amounts to
  $0.02\%$.  }

{The estimation is accurate even though we used only the
  measurements of the current flowing through the lines connecting
  $\tt{bus_{1}-bus_{39}}$, $\tt{bus_{9}-bus_{39}}$,
  $\tt{bus_{3}-bus_{4}}$, $\tt{bus_{16}-bus_{17}}$, and
  $\tt{bus_{14}-bus_{15}}$, (i.e., the branches indicated with dashed
  lines in \RefFig{F:IEEE39}), which are far away from the buses
  connecting the solar power plants.  As reported in the literature,
  these lines split the IEEE 39-bus system in areas.}

{The measurements used in this case highlight that we do not
  necessarily need to put \textsc{pmu}s in front of synchronous
  generators or grid-forming/following converters.  However, we do not
  have a systematic method that clearly indicates the optimal
  locations where \textsc{pmu}s should be connected to provide
  measurements used during the estimation process.  This is an open issue
  that we will cope with in future work.}

Moreover, we observe the birth of additional ``swing equation'' modes
due to the presence of this kind of the virtual synchronous generators
\cite{Hadavi2022}.  These, in fact, induce inter-area oscillations
that are clearly identified by peaks in the frequency responses by
frequency scan.  Modifying the inertia constant of these elements
changes significantly the variance of the measured electrical
quantities and, thus, helps increasing the accuracy of the proposed
technique.

\begin{figure}[htb]
  \centering
  \includegraphics[scale=1.0]{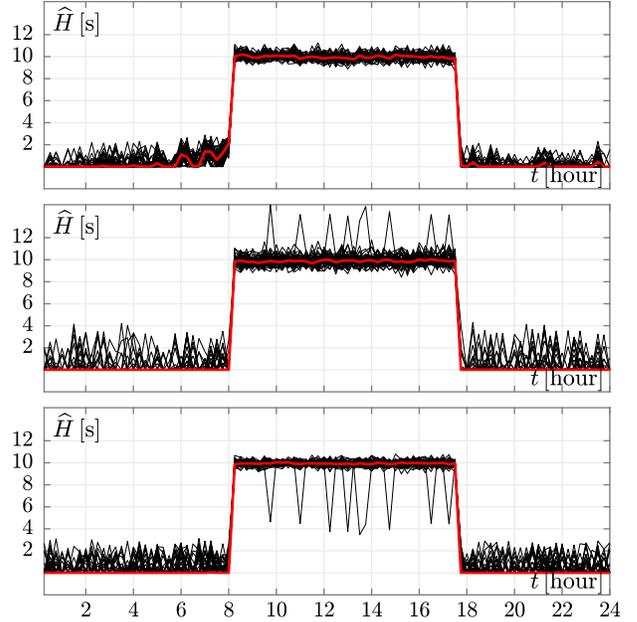}
  \caption{From top to bottom: inertia constant estimates over time of
    the grid-forming converters $\rm GFM_1$, $\rm GFM_2$, and
    $\rm GFM_3$ (all of which were set to $10\,\second$ from
    8:15~\amname{} to 5:30~\pmname{}). The results of the $50$
    estimation procedures are depicted in black. In red we report the
    median value computed every $15\,\minute$ over the $50$ trials.  }
  \label{F:SCEC_TRACES}
\end{figure}

Figure \ref{F:SCEC_TRACES} shows the estimation of the inertia of the
system across the full day of measurements.  This illustrates the
effectiveness of the proposed technique to {\em continuously} estimate
the inertia by stochastic fluctuations.  In Fig.~\ref{F:SCEC_TRACES},
the black traces are single trials while the red one is the median
value over $50$ trials.  All realisations give accurate results with
only a few time instants being affected by a relatively higher error.

\subsubsection{Inclusion of Grid-Following (GFL) Converters}
\label{ss:3}

In the third and last scenario, we discuss the effect of the frequency
droop control of grid-following converters on the estimation of the
inertia with the proposed technique.  As is well-known, the frequency
droop control of the converter is equivalent to a second order model,
and is thus similar to that of synchronous machines, except for the
fact that it shows a large damping and approximately zero inertia
\cite{6683080}. In this scenario, thus, we repeat the simulations of
the modified IEEE 39-bus system with inclusion of the same solar power
plants considered in the previous section. However, we assume that
these plants are now equipped with conventional frequency droop
controllers.

Table \ref{T:GFL} shows that the estimation of the damping/droop and
of the inertia are strongly correlated. An increase in the droop
coefficients of the frequency controllers leads to a lower inertia
constant estimation. This result had to be expected as a lower droop
makes the response of the controller slower and thus its time scale
tends to overlap with that of the inertial response. The estimation of
the inertia constant and of the damping coefficient are performed
independently, since two separate least-squares problems are solved:
one fits the inertia constant and a second one fits the damping
coefficient.

The main conclusion that can be drawn from the results shown in Table
\ref{T:GFL} is that the value of the inertia constant alone is not
sufficient to define the ability of a device to regulate the
frequency. If only the inertia constant is estimated, in fact, one
would conclude that the case with higher droop coefficient is the one
with lower frequency containment, which is not. On the other hand, if
both parameters, namely inertia and damping, are estimated, then one
can conclude, correctly, that the system with higher droop provides
``more'' fast frequency control than inertial response. On the other
hand, lower droop values increase the ability of the system to have an
inertial response but lead, as expected, to a lower capability of
providing a service as fast frequency controllers.

\begin{table}[t]
  \centering
  \caption{Estimation of the equivalent inertia and of the
    damping/droop for the grid-following converters in the modified
    IEEE 39-bus system.}
  \begin{tabular}{c|cc|cc}
    \toprule
    \multirow{2}{*}{Device} & \multicolumn{2}{c|}{$\text{droop}=2$} & \multicolumn{2}{c}{$\text{droop}=10$} \\
    % \cline{2-5}
    & $H_{eq}$ [$\second$] & $D_{eq}$ [$\mathrm{pu}$] & $H_{eq}$ [$\second$] & $D_{eq}$ [$\mathrm{pu}$] \\
    \midrule
    $\rm GFL_1$ & 0.64 & 3.29 & 0.40 & 8.65\\
    $\rm GFL_2$ & 0.60 & 3.13 & 0.37 & 7.97\\
    $\rm GFL_3$ & 0.56 & 3.93 & 0.34 & 10.57\\
    \bottomrule
    \end{tabular}
  \label{T:GFL}
\end{table}

\subsection{All-Island Irish Transmission System (\textsc{aiits})}
\label{S:AIITS}

{In this section, the proposed method is applied to a real-world
  complex system.  The model of the \textsc{aiits} considered here
  consists of 1479 buses, 1851 transmission lines or transformers, 245
  loads, 22 conventional synchronous power plants with \textsc{avr}s
  and turbine governors, 6 power system stabilizers, and 169 wind
  power plants.  Note that the secondary frequency control of the
  \textsc{aiits} is implemented manually and, thus, no \textsc{agc} is
  considered in the model.  Both load fluctuations and wind speeds are
  modeled as \ac{ou} processes.  The resulting set of DAEs for the
  \textsc{aiits} includes 2118 state variables (663 of which are
  stochastic processes) and 6175 algebraic variables.  Both the active
  and the reactive power of the $Z = 245$ loads is altered by an
  \ac{ou} process characterised by $v=0.1$, while $\sigma$ is defined
  in such a way that the standard deviation of those processes is
  $0.5\%$ of the nominal active or reactive power of each load.
  Finally, the standard deviation and drift term of the \ac{ou}
  processes modeling the wind speeds are set to $0.002$ and $0.02$,
  respectively.  The time domain simulations of the \textsc{aiits} are
  obtained with the software tool Dome \cite{6672387}.}

{ To test the applicability of the proposed method, we estimated
  the $\widehat H$ inertia constants of $11$ conventional synchronous
  power plants chosen to be representative of non-homogeneous $H$
  constants and power ratings (see Table \ref{T:PRATH2}).  An
  optimization problem was solved every $15\,\minute$ of simulated
  time by collecting $100$ estimates per synchronous generator. Making
  reference to the diagram in Figure \ref{F:DIAGRAM}, the subset
  $\mathcal Z$ contains only the magnitude of the voltage at the buses
  where the generators are connected. Hence we did not resort to the
  phase of such voltages, to any current of the network, or to any
  rotor speed of the synchronous generators.}

\begin{figure}[t]
\centering
\includegraphics[scale=1.0]{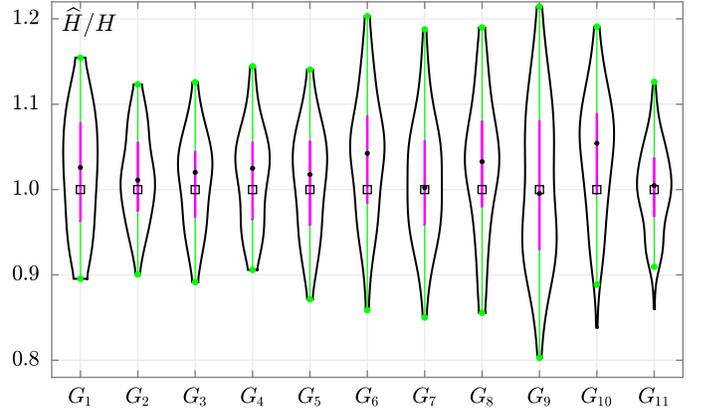}
\caption{{Violin plots of the inertia constants estimates for
    some generators of the \textsc{aiits}, normalized w.r.t. their
    effective value.}
\label{F:AIITS}}
\end{figure}

{The $\epsilon_{\%}^H $ relative percentage error spans from
  $0.22\%$ (for $G_7$) to $5.42\%$ (for $G_{10}$). The quality of the
  results, measured through $\epsilon_{\%}^H$, is similar to that
  obtained for the IEEE 39-bus system.}  { For the sake of
  readability, the violin plots in \RefFig{F:AIITS} report the
  statistics of the $\widehat H$ estimated inertia values normalized
  w.r.t. the $H$ nominal inertia of each generator. So doing, for each
  violin plot in \RefFig{F:AIITS} the reference value is $1$. The
  distance of the upper and lower adjacent values (green solid dots in
  Figure \ref{F:AIITS}) from the expected normalized value of
  $\widehat H$ is lower than $0.22$ for all the $11$ synchronous
  generators, which confirms the robusteness of the proposed method
  when applied to a real-world complex system with several hundreds of
  buses and dynamic devices.}

\begin{table}[htb]
  \caption{{Synchronous generators $H$ and $P_{\mathrm{RAT}}$ of the \textsc{aiits}}}
  \centering
    \begin{tabular}{ c | r | c | c | r | c}
      \toprule
      Gen. & $H\,[\second]$& $P_{\mathrm{RAT}}\,[\mega\watt]$ &
      Gen.	& $H\,[\second]$& $P_{\mathrm{RAT}}\,[\mega\watt]$\\
      \midrule
      $G_{1}$ & $15.924$  & $535.5$ & $G_{7}$ & $10.000$ & $135$ \\
      $G_{2}$ & $13.400$  & $288$ & $G_{8}$ & $11.652$ & $292$\\
      $G_{3}$ & $7.320$  & $87.5$ & $G_{9}$ & $7.744$ & $353$\\
      $G_{4}$& $7.320$& $87.5$& $G_{10}$& $7.744$& $127.6$\\
      $G_{5}$& $7.320$& $87.5$& $G_{11}$& $10.700$& $359$\\
			$G_{6}$& $14.344$& $22.47$& $ $& $ $& $ $\\
	\bottomrule
    \end{tabular}
    \label{T:PRATH2}
\end{table}

\section{Conclusions}
\label{S:Con}

This work proposes a technique to estimate the inertia based on the
variance of measurements.  Assuming a given model of the power system
and the dynamics of the generators, the technique fits this model to
the measurements by solving a least-squares problem.  Simulation
results show that the proposed technique, given an adequate filtering
of the measurements, is accurate {even in a real-world complex
  system} and can take into account both conventional and virtual
synchronous machines, as well as the effect of frequency droop
controllers. However, the correct interpretation of the effect of
droop controllers is consistent only if also the damping/droop
coefficients are estimated, which the proposed technique duly allows
to obtain.

In particular, we believe that the proposed technique can be useful to
reward the frequency support provided by non-synchronous devices. In
turn, we recommend that the reward should be evaluated based on the
\textit{effect} that the controllers of these devices have on the
system, not on the actual implementation of the control itself.

{Future work will focus on further investigating the theoretical
  and practical aspects of the proposed technique.  For example,
  relevant questions are how to improve the accuracy of the estimated
  inertia constants through (i) an optimal number and location of the
  set of measurements and (ii) of the filtering of the measurements.}

% In particular, the first point involves the attainment of a strategy
% to identify the number of \textsc{pmu}s, as well as their optimal
% location in the network.

% ----- ----- ----- ----- ----- ----- ----- ----- -----
%\bibliographystyle{IEEEtran}
%\bibliography{biblio}

% ----- ----- ----- ----- ----- ----- ----- ----- -----

\begin{IEEEbiography}[{\includegraphics[width=1in,height=1.25in,clip,
    keepaspectratio]{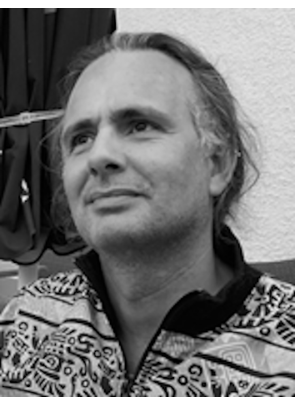}}] {Federico
    Bizzarri}(M'12--SM'14) was born in Genoa, Italy, in 1974.  He
  received the Laurea (M.Sc.)  five-year degree (\textit{summa cum
    laude}) in electronic engineering and the Ph.D. degree in
  electrical engineering from the University of Genoa, Genoa, Italy,
  in 1998 and 2001, respectively.  Since October 2018 he has been an
  associate professor at the Electronic and Information Department of
  the Politecnico di Milano, Milan, Italy.  He is a research fellow of
  the Advanced Research Center on Electronic Systems for Information
  and Communication Technologies ``E. De Castro'' (ARCES), University
  of Bologna, Italy.  He served as an Associate Editor of the IEEE
  Transactions on Circuits and Systems --- Part I from 2012 to 2015
  and he was awarded as one of the 2012-2013 Best Associate Editors of
  this journal.
\end{IEEEbiography}

\begin{IEEEbiography}[{\includegraphics[width=1in,height=1.25in,clip,
    keepaspectratio] {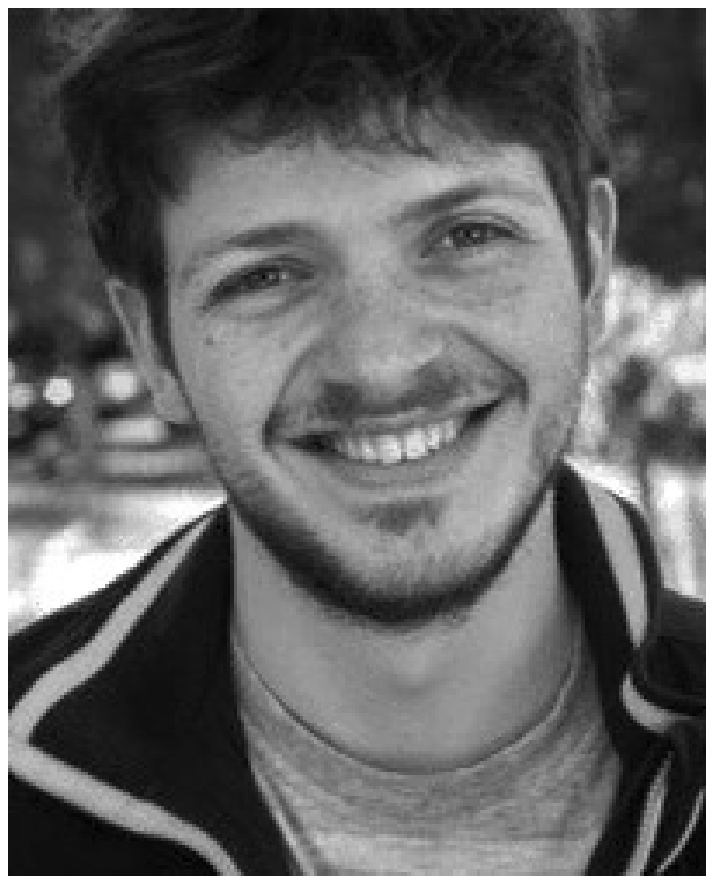}}] {Davide del Giudice} was born in
  Milan, Italy, in 1993. He received the M.S. degree and the
  Ph.D. degree in electrical engineering from Polytechnic
  of Milan in 2017 and 2022, respectively. He is currently a
  researcher at the department of Electronics, Information and
  Bioengineering of Polytechnic of Milan. His main research
  activities are related to simulation techniques for
  electric power systems with a high
  penetration of converter-interfaced elements, such as high voltage
  direct current systems, electric vehicles and generation fuelled by
  renewables.
\end{IEEEbiography}

\begin{IEEEbiography}[{\includegraphics[width=1in,height=1.25in,clip,
keepaspectratio]{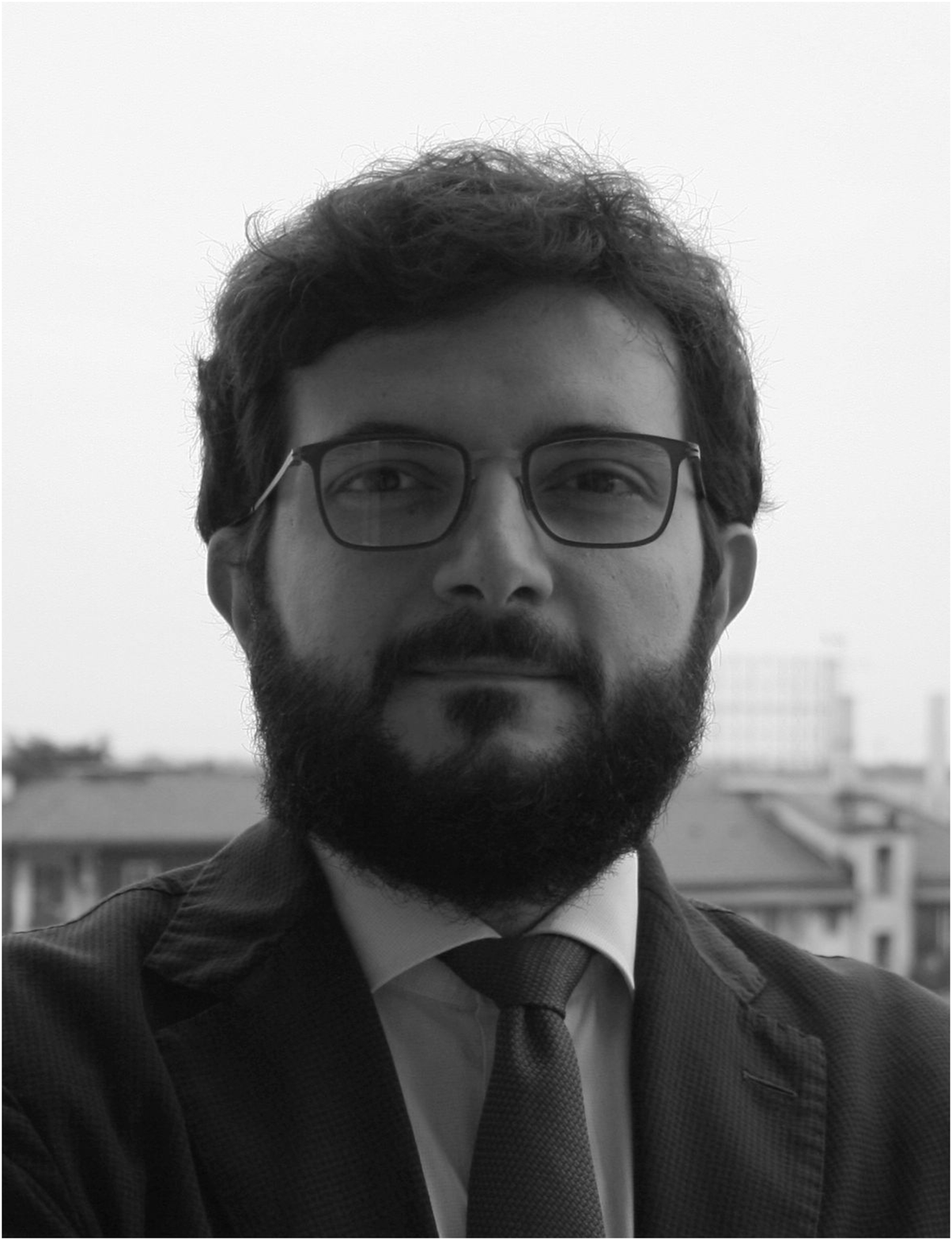}}]{Samuele Grillo} %
(S'05--M'09--SM'15) received the Laurea degree in electronic engineering, and the Ph.D. degree %
in power systems from the University of Genova, Italy, in 2004 and 2008, respectively. %
He is currently an Associate Professor with the Dipartimento di Elettronica, Informazione e %
Bioingegneria, Politecnico di Milano, Milan, Italy.

Since 2018 he is a contributor to CIGR\'{E} Working Group B5.65 ``Enhancing Protection System %
Performance by Optimising the Response of Inverter-Based Sources.''

His research interests include smart grids, integration of distributed renewable sources, %
and energy storage devices in power networks, optimization, and control techniques applied %
to power systems.
\end{IEEEbiography}

\begin{IEEEbiography}[{\includegraphics[width=1in,height=1.25in,clip,
    keepaspectratio]{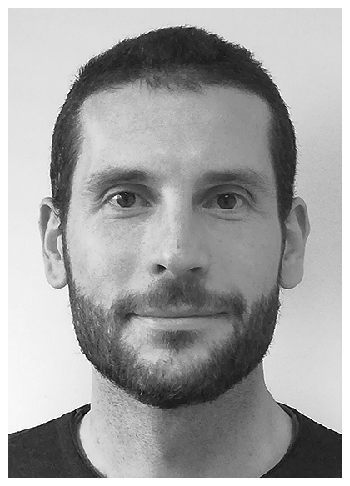}}]{Daniele Linaro} Daniele
  Linaro received his MSc in Electronic Engineering from the
  University of Genoa (Italy) in 2007 and a PhD in Electrical
  Engineering from the same university in 2011. Since 2018, he is
  Assistant Professor in the Department of Electronics, Information
  Technology and Bioengineering at the Polytechnic of Milan. His main
  research interests are currently in the area of circuit theory and
  nonlinear dynamical systems, with applications to electronic
  oscillators and power systems and computational neuroscience, in
  particular biophysically-realistic single-cell models of neuronal
  cells.
\end{IEEEbiography}

\begin{IEEEbiography}[{\includegraphics[width=1in,height=1.25in,clip,
    keepaspectratio] {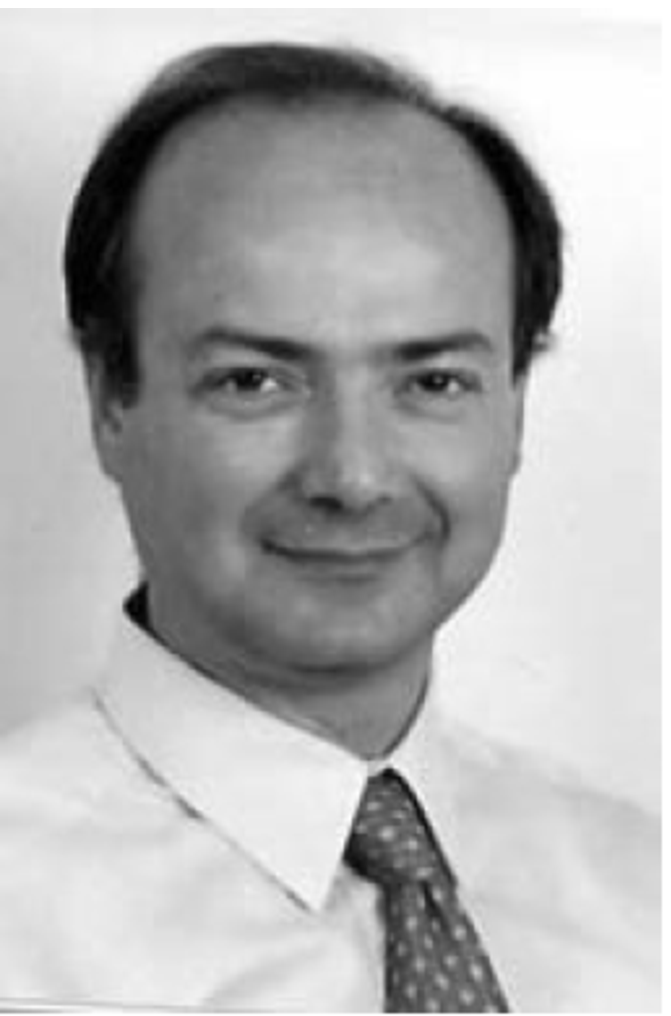}}] {Angelo Brambilla}
  (M'16) received the Dr.~Ing.~degree in electronics engineering from
  the University of Pavia, Pavia, Italy, in 1986. He is full professor
  at the Dipartimento di Elettronica, Informazione e Bioingegneria,
  Politecnico di Milano, Milano, Italy, where he has been working in
  the areas of circuit analysis, simulation and modeling.
\end{IEEEbiography}

\begin{IEEEbiography}[{\includegraphics[width=1in, height=1.25in,
    clip, keepaspectratio]{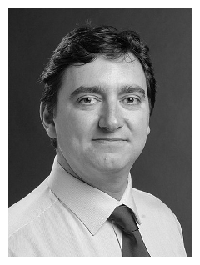}}]{Federico Milano} (F'16)
  received from the Univ.~of Genoa, Italy, the ME and Ph.D.~in
  Electrical Eng.~in 1999 and 2003, respectively.  From 2001 to 2002
  he was with the Univ.~of Waterloo, Canada, as a Visiting Scholar.
  From 2003 to 2013, he was with the Univ.~of Castilla-La Mancha,
  Spain.  In 2013, he joined the Univ.~College Dublin, Ireland, where
  he is currently Prof.~of Power Systems Control and Protections and
  Head of Elec.~Eng.  In 2022, he was appointed Co-Editor in Chief of
  the IET Generation Transmission \& Distribution.  His research
  interests include power system modeling, control and stability
  analysis.
\end{IEEEbiography}

\end{document}